\documentstyle{article}
\begin{document}
\title{Statistical properties of the faint
early-type galaxies sample: possible effects of evolution}
\author{I.A.Trifalenkov}
\maketitle
\begin{center}
\it
Space Research Institute, Profsojuznaja 84/32, Moscow, 117810, Russia
\newline
E-mail: itrifaln@esoc1.iki.rssi.ru
\end{center}

\begin{abstract}
Early-type galaxies from CfA catalogue were assotiated with IRAS FSC
sources. In given sample IR and optical properties were used to obtain
some classification of these objects. The existance of significant classes
can be evidence of presence one main type of the dust in each galaxy.
Physical interpretation of obtained classes was discussed. The dust mass
in two classes of galaxies can be calculated using evolution modelling.
In this case we can obtain a estimations for chemical composition and
initial mass function in these galaxies.
\end{abstract}
\newpage
\section{Introduction}
The availability of an enormous amount of infrared
data, obtained in the IRAS experiment, makes it possible to
analyze statistical properties of samples of various celestial
objects, including those in which the components of IR emission
are expected to be fairly weak. In particular, galaxies of early
morphological types (E and S0) may be placed into the latter
category. Although the first release of the IRAS Point Source
Catalog contained only a small number of such galaxies, the
application of additional processing techniques allowed the
formation of a large sample of such objects and the
investigation of statictical properties of various galaxy
physical parameters, obtained both from IR data and from data
derived in other spectral regions (Knapp et al. \cite{knapp:all}, Bally \&
Thronson \cite{bally}).
The release of the next catalog of IR
sources (Faint Source catalogue v.2, FSC-2, Moshir et al \cite{mosh:all})
justifies the
studies of general sample properties of early-type galaxies,
associated with the sources of this catalog. These studies are
also useful because the objects in the IRAS Point Source Catalog
(PSC) and FSC-2 Catalog are subject to various selection effects
and may exhibit various physical properties. In this work, such
an investigation was performed using objects from the CfA
catalog (Huhra, \cite{huhra}), which has associations in the FCS-2
catalog. For the derived sample, the objects were classified by
means of the cluster analysis, and physical interpretation of
the obtained classes was made.

\section{Data and data handling}
To analyze possible associations with
Faint Source Catalog, the largest available catalog of galaxies
- CfA catalog - was used, which containes 38909 objects;
morphological type is given for 17008 objects, 4596 objects from
them are early-type galaxies (catalog's type $t < 0$). To apply
the identification procedure, the following selection criteria
were used for objects: the separation between the objects in the
CfA and FSC-2 catalogs was within 60 arcsec; the object showed a
statistically significant flux (parameter ${FQUAL} > 2$) in one of
the far-IR bands (60 or 100 $\mu m$); there was no associations for
the source in FSC-2 catalog with other, closer objects from
catalogs of other objects. The applied procedure resulted in a
sample of 539 galaxies, which was used in the subsequent
analysis; 11 galaxies from this list had no previous
associations with IR sources.  A group of galaxies with
statistically significant IR measurements in, at least, three
IRAS bands and with measured apparent magnitudes in Johnson's
UBV system, as well as with known redshifts, was selected from
the general sample. These data were available for 166 E and S0
galaxies. Such selection is required to classify objects by, at
least, three parameters. As it is desirable that the selected
parameters were independent of the galaxy distance, the flux
ratios at 25, 60, 100 $\mu m$, and in the $B$ band ($f_{25}/f_{60}$,
$f_{100}/f_{60}$,
$f_B/f_{60}$) were used for the classification.

Hierarchical
clustering algorithm, involving a sequential agglomeration of
objects or a group of objects (clusters) using a criterion of
any kind, was applied for object classification. After the
agglomeration of objects into a single cluster, the maximum
number of statistically significant clusters is determined using
a criterion of any kind (a set of criteria). In this case, a
software program of hierarchical cluster analysis, taken from
the STATLIB library (compiled by Murtagh \cite{murt})
minimal increase in residual mean square (RMS) in a cluster during
 agglomeration
(Ward's criterion) was used as a criterion for cluster
agglomeration; this leads to breaking the hyperspherical form
(in classification parameter space) into clusters. To estimate
the significance of breaking into clusters, a semi-empirical
criterion was applied, based on an analysis of the
"dissimilarity - number of cluster" relationship
(Mojena \& Wishart \cite{mojena:wish});
according to this criterion, the agglomeration
must stop when the growth rate of RMS increases during cluster
agglomeration. The requirement that the separation between
clusters be larger than  RMS within each cluster was used as an
additional criterion. The application of both these criteria
resulted in breaking the sample into four clusters (object
classes), which were subsequently subject to further analysis.

\section{Results of Data Processing}
To analyze the derived classes of
galaxies, some additional physical parameters for each sample
object were calculated, including the dust distribution in
temperature, dust masses, and the contribution of IR emission to
overall emission from galaxies. The limited number of
photometric bands require a simple dust model to describe the
observational IR photometry:
$$
I_{\nu} = \tau_{\nu} B_{\nu}\left(T_{\mbox {dust}}\right)
$$
where $\tau_{\nu} \propto \nu^{\alpha}$ is the optical depth,
and $B_{\nu}\left(T_{\mbox {dust}}\right)$ - the Plank
function. Unfortunately this approximation can not give
meaningful results for a large number of galaxies in the sample.
As a consequence, we have to use a multicomponent dust model:
$$
I_{\nu} = \sum_{i=1}^n\tau_{\nu}^{(i)} B_{\nu}\left({T_{\mbox
{dust}}^{(i)}}\right)
$$
A two-component ($n=2$) model of dust was employed to evaluate the
temperature distribution. As three measurements are not enough
to determine the complete set of parameters for such a model,
the spectral index $\alpha$ in the dependence of absorption coefficient
on the frequency and the temperature of one of the
component ($T_{\mbox dust}^{(1)} = 28 {\mbox K}$) were fixed parameters.
As a result of
the modelling, the temperature of the "hot" component, as well
as the "hot"- to-"cold" component ratio was calculated. The
"hot" component with temperature $T_{\mbox dust}^{(2)} > 60 {\mbox K}$
is present
virtually in all galaxies, its contribution to overall emission
being in the range from $1\%$ to $60\%$. The mass of dust, as in the
paper by Greenhouse et al \cite{ght}, was
calculated from the emission in a wavelength interval of 25 to
300 $\mu m$, taking into account the two-component model of dust.
Statistical data on classification parameters ($lg(f_B/f_{60})$,
$lg(f_{25}/f_{60})$, $lg(f_{100}/f_{60})$) and additional physical parameters,
quoted above ($T_{\mbox dust}^{(2)}, lg(M_{\mbox dust})$, the contribution
of "hot"
component to overall emission , as well as the redshift $z$ and
morphological type $t$), were obtained for each selected class of
galaxies and are given in Table 1.

\begin{tabular}{|c||l||c||c||c|}
\hline
Parameter & Cluster1 & Cluster2 & Cluster3 & Cluster4\\
\hline
\hline
Redshift $z$ & 0.017 $\pm$ 0.001 & 0.010 $\pm$ 0.003 & 0.019 $\pm$ 0.003 &
0.009 $\pm$ 0.004 \\
Type $t$  & -2.86 $\pm$ 0.21  & -2.32 $\pm$ 0.25  & -2.56 $\pm$ 0.26  & -2.78
$\pm$ 0.35  \\
$\lg (f_B)/f_{60})$ & -0.36 $\pm$ 0.09  &  1.34 $\pm$ 0.13  &  0.49 $\pm$ 0.11
&  3.35 $\pm$ 0.21 \\
$\lg (f_{25})/f_{60})$ & -0.86 $\pm$ 0.01 & -0.99 $\pm$ 0.02 & -0.49 $\pm$ 0.03
& -0.49 $\pm$ 0.03 \\
$\lg (f_{100})/f_{60})$ & 0.23 $\pm$ 0.01  & 0.41 $\pm$ 0.01  & 0.20 $\pm$ 0.02
 & 0.53 $\pm$ 0.03 \\
$T_{\rm dust}^{(2)}$ & 66.41 $\pm$ 0.41 & 69.89 $\pm$ 2.02 & 81.54 $\pm$ 1.50 &
94.49 $\pm$ 4.81 \\
$\lg (M_{\rm dust})$ in $M_{\odot}$ & 7.60 $\pm$ 0.09 & 7.15 $\pm$ 0.15 & 7.52
$\pm$ 0.12 & 6.51 $\pm$ 0.21 \\
"Hot"/ "Total" & 0.37 $\pm$ 0.02 & 0.27 $\pm$ 0.05 & 0.38 $\pm$ 0.03 & 0.13
$\pm$ 0.02 \\
\hline
\hline
\end{tabular}

To interpret some physical
properties of galaxies of various types, the available
observations revealing traces of interaction (possible
associations with objects from Vorontsov-Vel'aminov's catalod of
interacting galaxies), traces of an active nucleus (possible
associations with objects from Veron's catalog), and the
presence of radio emission, were collected. As a result of the
comparison of the above signs, the following interpretation may
be provided for the division of early-type galaxy sample into
classes:
\begin{itemize}
\item Class 1: Galaxies with high abundance of
moderate-temperature dust; this class includes all interacting
galaxies, galaxies with giant HII zones, and a number of sources
emitting in a high-frequency radio range; the above signs imply
outbreaks of star formation in this class of galaxies and the
existence of UV photons - the main source of dust heating.
\item Class 2: Differs from the previous class by a considerably lower
dust abundance and by a minimum number of peculiar properties
(there are weak traces of nuclear activity in a number of
cases); this class appears to include "normal" ellipticals.
\item Class 3: Contains mainly galaxies with active nuclei,
high-temperature dust (when the amount of optical emission,
reradiated in infrared, is small), galaxies with synchrotron
radio sources. The dust properties being close to IR properties
of galaxies implies that the dust surrounding the nucleus is the
main contributor of IR emission in this class of galaxies.
\item Class 4: This class of objects has the same properties as
class-2 galaxies, except that their dust is warmer. To establish
the nature of the additional dust-heating source for this class
of galaxies a more thorough investigation of their observed
properties in a variety of spectral regions may be required.
Another explanation can be offered, if we assume that the
spectral index $\alpha$ in the dust model is large. Since $\alpha$ is
determined by the size distribution of dust grains, the
distruction of large dust particles takes place in these
objects. The negative corelation between the temperature and
mass of the dust favors the second explanation.
\end{itemize}

\section{Dust and history of galaxies}
As a result previous consideration, we can see, that a set of galaxies
without significant star formation, nuclear activity and interaction can
be selected using by clustering methods. For these objects we can suggest
late-type stars in each galaxy as ultimate source of interstellar dust
in this galaxy. It provide us to write relations between stellar evolution
and dust mass in a galaxy. If the galactic mass is generated mainly  in
atmospheres of late stars, dust mass can be estimated as:
$$
M_{\rm dust} = \zeta(Z) \: t_{\rm dust} \: {\mbox{\.{M}}_{\rm gas}}
$$
where $\zeta(Z)$ is the content of dust in the stellar wind (depend on
chemical composition $Z$), $t_{\rm dust}$ -- lifetime of dust grains,
$\mbox{\.{M}}_{\rm gas}$ - stellar mass loss rates, determined as:
$$
\frac{\mbox{\.{M}}_{\rm gas}}{M_{\rm gal}} = \int_{M_{\rm min}}^{M_{\rm max}}
{\frac{{\mbox{\.{M}}}(M_*)}{M_*}} {\phi(M_*)} dM_*
$$
where $M_{\rm gal}$ is total galactic mass, $M_{\rm min}$, $M_{\rm max}$
are minimal and maximal stellar masses, and $\phi(M_*) \propto {M_*}^{-\alpha}$
--
stellar mass function. Since $\zeta(Z)$ and $\mbox{\.{M}}(M_*)$ can be obtained
from stellar models, dust mass, obtained from IR-observations can be useful
tool to estimate the chemical composition and stellar mass function in the
galaxies.
\section{Conclusions}
\begin{enumerate}
\item It is possible to obtain statistically significant
groups of galaxies on the basis of IRAS photometry. This fact
can be considered as evidence of the existence of one main type
of dust in each galaxy.

\item The derived classes of galaxies can be interpreted as: -
galaxies with active star formation (class 1); - galaxies with
strong nuclear activity (class 3); - "normal" E/S0 galaxies
(class 2); - galaxies with large index in the size distribution
of grains (class 4).

\item Many galaxies show appreciable (up to 80\% of the overall dust
mass) amount of hot ($T_{\mbox dust} > 60 {\mbox K}$) dust.

\item For galaxies from classes 2 and 4 we have an opportunity to estimate
the chemical composition and initial mass function from the dust mass.

\end{enumerate}

\end{document}